\documentclass[letterpaper, 10 pt, conference]{ieeeconf}  

\IEEEoverridecommandlockouts                              

\usepackage{amsmath,amsfonts,amsthm}
\usepackage{array}
\usepackage[caption=false,font=normalsize,labelfont=sf,textfont=sf]{subfig}
\usepackage{graphicx}
\usepackage{cite}
\usepackage[short]{optidef}
\usepackage{longtable} 
\usepackage{booktabs}
\usepackage{adjustbox}
\usepackage[dvipsnames]{xcolor}
\usepackage{rotating}
\usepackage{pifont}
\newcommand{\xmark}{\ding{55}}%

\newtheorem{definition}{Definition}[section]

\newtheorem{lemma}{Lemma}[section]
\newtheorem{corollary}{Corollary}[section]

\usepackage{acronym}
\newacro{QP}{quadratic program}
\newacro{MIQP}{mixed-integer quadratic program}
\newacro{MINLP}{mixed-integer nonlinear program}
\newacro{MIOCP}{mixed-integer optimal control problems}
\newacro{MI}{mixed-integer}
\newacro{MIP}{mixed-integer program}
\newacro{NLP}{nonlinear program}
\newacro{NMPC}{nonlinear model predictive control}
\newacro{FF}{feed forward}
\newacro{DNN}{deep neural network}
\newacro{FCF}{Frenet coordinate frame}
\newacro{CCF}{Cartesian coordinate frame}
\newacro{FP}{feasibility projector}
\newacro{SQP}{sequential quadratic programming}
\newacro{RL}{reinforcement learning}
\newacro{AD}{autonomous driving}
\newacro{RTI}{real-time iteration}
\newacro{ODE}{ordinary differential equation}
\newacro{SV}{surrounding vehicle}
\newacro{EV}{ego vehicle}

\newcommand{\opp}{\ac{SV}}
\newcommand{\opps}{\acp{SV}}
\newcommand{\rti}{\ac{RTI}}

\newcommand{\nmpc}{\ac{NMPC}}

\newcommand{\fcf}{\ac{FCF}}

\newcommand{\pnorm}{\emph{ScaledNorm}}
\newcommand{\logsumexp}{\emph{LogSumExp}}
\newcommand{\boltzmann}{\emph{Boltzmann}}
\newcommand{\panos}{$\mathrm{ReLU}^2$}

\newcommand{\refer}[1]{\tilde #1}
\newcommand{\ub}[1]{\overline{#1}}
\newcommand{\lb}[1]{\underline{#1}}
\newcommand{\frenetTrans}{\mathcal{F_{\tilde{\gamma}}}}

\newcommand{\td}{t_\mathrm{d}}

\newcommand{\dimT}{t_\mathrm{f}}

\newcommand{\dimHor}{N}

\newcommand{\selMatX}{P_p}
\newcommand{\selMatV}{P_v}

\newcommand{\ball}{\mathcal{B}}

\newcommand{\setEgoCirc}{\mathcal{O}^\mathrm{ego}}

\newcommand{\setOpp}{\mathcal{O}^\mathrm{sv}}
\newcommand{\setOppExact}{\mathcal{O}^{\mathrm{sv}*}}
\newcommand{\setOppEll}{\mathcal{O}^\mathrm{ell}}

\newcommand{\setOppPnorm}{\mathcal{O}^\mathrm{p}}
\newcommand{\setOppLog}{\mathcal{O}^\mathrm{lse}}
\newcommand{\setOppBoltz}{\mathcal{O}^\mathrm{bm}}

\newcommand{\setFreeOppExcat}{\mathcal{F}^{\mathrm{sv}*}}

\newcommand{\freeSpace}{\mathcal{F}}

\newcommand{\stateAdmiss}{\mathcal{X}}
\newcommand{\conrlAdmiss}{\mathcal{U}}

\newcommand{\obsFun}{o}

\newcommand{\R}{\mathbb{R}}
\newcommand{\Z}{\mathbb{Z}}

\newcommand{\T}{^\top}
\newcommand{\norm}[1]{\left\lVert #1 \right\rVert}

\newcommand{\gen}{\xi}

\title{\LARGE \bf
Progressive Smoothing for Motion Planning in Real-Time NMPC
}

\author{Rudolf Reiter$^{1}$, Katrin Baumgärtner$^{1}$, Rien Quirynen$^{2}$ and Moritz Diehl$^{1,3}$
\thanks{$^{1}$Department of Microsystems Engineering, University Freiburg, 79110 Freiburg, Germany
	{\tt\small \{rudolf.reiter, katrin.baumgaertner, moritz.diehl\}@imtek.uni-freiburg.de}}%
\thanks{$^{2}$Mitsubishi Electric Research Laboratories, Cambridge, MA, USA \tt\small quirynen@merl.com}
\thanks{$^{3}$Department of Mathematics, University Freiburg, 79110 Freiburg, Germany
}%
}%

\begin{document}

\maketitle
\thispagestyle{empty}
\pagestyle{empty}

\begin{abstract}
\Ac{NMPC} is a popular strategy for solving motion planning problems, including obstacle avoidance constraints, in autonomous driving applications.
Non-smooth obstacle shapes, such as rectangles, introduce additional local minima in the underlying optimization problem. Smooth over-approximations, e.g., ellipsoidal shapes, limit the performance due to their conservativeness. 
We propose to vary the smoothness and the related over-approximation by a homotopy. Instead of varying the smoothness in consecutive \ac{SQP} iterations, we use formulations that decrease the smooth over-approximation from the end towards the beginning of the prediction horizon. Thus, the \rti{} algorithm is applicable to the proposed \ac{NMPC} formulation. Different formulations are compared in simulation experiments and shown to successfully improve performance indicators without increasing the computation time.
\end{abstract}

\acresetall 
\section{Introduction}
\label{sec:introduction}
Motion planning problems with obstacle avoidance are efficiently solved by derivative-based nonlinear optimization algorithms such as \ac{SQP}~\cite{Diehl2005, Rosolia2018,Kloeser2020,Varquez2020ab,Xing2022,Ayoub2022}.
These algorithms pose limitations on the problem formulation to have beneficial numerical properties. 
Among others, a major desired property is smoothness in the constraints. Particularly for obstacles, which in \ac{AD} applications are \opps{}, it was shown that ellipsoids achieve superior performance compared to other formulations~\cite{Reiter2023a}. 
However, ellipsoids over-approximate the supposed rectangular \opp{} shape by a large extent. Tighter formulations, such as higher-order norms can more accurately represent the shape, given an initial guess sufficiently close to an optimum~\cite{Bergman2018,Schimpe2020}. However, rectangular and higher-order ellipsoidal \opp{} shapes are prone to introduce local minima and linearizations are discontinuous due to the obstacle corners. Thus, numerical solvers may get stuck in local minima, as shown empirically in experiments. 

\begin{figure}
	\centering
	\includegraphics[scale=0.65]{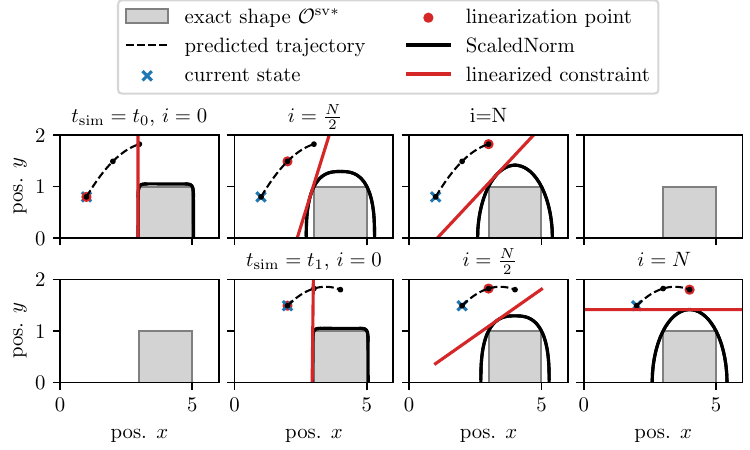}
	\caption{Progressive smoothing with the proposed \pnorm{} of the obstacle shape along the prediction steps~$i\in\{0,N/2,N\}$ for an NMPC prediction of $N$~steps at simulation 
time~$t_\mathrm{sim}=t_0$ (first row) and $t_\mathrm{sim}=t_1$ (second row). The \pnorm{} is smoothest for~$i=N$ and tightest for~$i=0$. Each plot shows the linearization of the \pnorm{} at the related prediction step.}
	\label{fig:lin_prog}
\end{figure}

The presented idea builds on the assumption that predictions at the end of the horizon are more uncertain and planned motions can more easily be adapted. This flexibility makes the accurate \opp{} shape less important which we exploit to represent the \opp{} with more favorable numerical properties near the end of the horizon. The shape is transformed to a more non-smooth one towards the beginning of the horizon, cf., Fig.~\ref{fig:lin_prog}, where the uncertainty and flexibility is lower and also the receding horizon shifting of the previous solution provides a good warm start to a desirable local minimum. The loss of optimality due to constraint over-approximation can be reduced significantly if the over-approximations are rather tight near the beginning of the horizon. We refer to the proposed shape transformation as~\emph{progressive smoothing}.

The novel approach, referred to as \pnorm{}, builds on \nmpc{} in the Frenet coordinate frame~\cite{Reiter2023a}, which is summarized in Sec.~\ref{sec:problem_setting}.
The \pnorm{} and two alternative formulations, referred to as \logsumexp{} and \boltzmann{}, are introduced and analyzed among their essential numerical properties for obstacle avoidance in Sec.~\ref{sec:method}.
Within closed-loop simulations, the performance is compared in Sec.~\ref{sec:experiments}, including the overtaking distance, the susceptibility for getting stuck in local minima and the computation time.

\subsection{Related Work}
\label{sec:related}
An abundance of authors have successfully applied \nmpc{} for \ac{AD} with obstacle avoidance~\cite{Sathya2018,Rosolia2018,Kloeser2020,Varquez2020ab,Xing2022,Ayoub2022}, and an accurate representation of obstacle shapes is a major concern.

Due to favorable numerical properties, over-approximating ellipses are used in~\cite{Geisert2016,Nair2022}. 
The authors in \cite{Bergman2018, Schimpe2020, SCHOELS2020} introduce higher order norms and~\cite{Oh2021} uses the infinity norm. However, as shown within this work, they are susceptible of getting stuck in local minima.
Several covering circles~\cite{Ziegler2014a,Jinxiang2021}, a smooth infinity norm approximation~\cite{Sathya2018} referred to as~\panos{}, or separating hyperplanes~\cite{Brossette2017, Nair2022} are used to more accurately capture the shape, however, as shown in~\cite{Reiter2023a}, the performance is worse compared to ellipses on a number of test examples.

The authors in~\cite{Bergman2018} use homotopies combined with~\ac{SQP} iteration to improve convergence by smoothing obstacle shapes.
Progressive \emph{smoothing} refers to the idea of progressively smoothing and expanding obstacle shapes along the horizon. Equally, this can be formulated as progressively \emph{tightening} the feasible set along the horizon as it was introduced in~\cite{Zanelli2021c, Baumgaertner2023}. In~\cite{Zanelli2021c}, the authors reformulate constraints as costs on a part of the horizon for reduced computational complexity. Despite originating from a different idea, asymptotic stability for a general class was shown for this tightening in~\cite{Baumgaertner2023}. 

\subsection{Contributions}
\label{sec:contributions}
We contribute a novel formulation for progressive smoothing of obstacle shapes and a numerical analysis with respect to obstacle avoidance for \nmpc{}.
In addition we introduce two alternative formulations for progressive smoothing.
Their performance in closed-loop simulations is evaluated and highlighted against other state-of-the-art formulations.
\subsection{Preliminaries}
\label{sec:prelim}
For index sets the notation $\mathbb{I}(n)=\{0,1,\ldots,n\}$ is used. Furthermore $\Z^+$ refers to the strictly positive integer numbers. 
With the operator $[x]_i$, the $i$-th element of the vector $x\in\R^n$ is selected and the expression $\lfloor x \rfloor$ is used to denote the $\mathrm{floor}(\cdot)$ function, i.e., rounding down.
\section{Problem Setting}
\label{sec:problem_setting}
The problem of motion planning and control of an autonomous vehicle is considered. 
Particularly, a vehicle is controlled in a structured road environment that either involves driving along a certain reference lane~\cite{Quirynen2023} or approximating time-optimal driving for racing applications~\cite{Kloeser2020, Reiter2021, Reiter2023b} while avoiding obstacles of a rectangular shape.

As shown in several works~\cite{Werling2010a,Rosolia2018,Kloeser2020,Varquez2020ab,Xing2022,Reiter2023a}, using \nmpc{} with a model formulation in the \fcf{} yields state-of-the-art performance. Notably, the presented method is independent of the coordinate frame chosen for the model representation. However, the model representation is essential for the \nmpc{} formulation.
In the following, the basic concept of the \fcf{} \nmpc{} is defined.

The \fcf{} transformation projects Cartesian position states~$p^\mathrm{veh}\in\R^2$, together with a vehicle heading angle~$\phi\in\R$ to a curvilinear coordinate system along a curve~$\gamma(s):\R \rightarrow \R^2$, with the path position~$s\in\R$.
The Cartesian states~$x^{\mathrm{c}}=[p^\mathrm{veh},\;\phi]^{\T}$ are transformed to the \fcf{} states~$x^\mathrm{f}=[s,\;n,\;\beta]^\top$, with the longitudinal road aligned position~$s$, the lateral position~$n$ and the heading angle mismatch~$\beta$. The closest point on the reference curve~$s^*$,
\begin{equation}
	\label{eq:frenet_transformation_argmin}
	s^*(p^\mathrm{veh})=\arg \min_{\sigma}\norm{p^\mathrm{veh}-\gamma(\sigma)}^2_2,
\end{equation}
is used for the Frenet transformation, defined as
\begin{align}
	\label{eq:frenet_transformation}
	x^\mathrm{f}=\frenetTrans(x^\mathrm{c})=\begin{bmatrix}
		s^* \\
		(p^\mathrm{veh}-\gamma(s^*))^\top e_n(s^*) \\
		\phi^\gamma(s^*)-\phi 
	\end{bmatrix},
\end{align}
where $\phi^\gamma(s)$ is the tangential angle of the curve~$\gamma(s)$ and $e_n(s):\R\rightarrow\R^2$ is the normal unit vector to the curve.

We use a \fcf{} kinematic vehicle model with~$n_x=5$ states $x=[s,\; n,\; \beta,\; v,\;\delta]^\top\in\R^{n_x}$ and a wheelbase~$l$. It includes the steering angle~$\delta$ and the velocity~$v$, the inputs~$u=[F^\mathrm{d},\;r]$ with the longitudinal acceleration force~$F^\mathrm{d}$ and the steering rate~$r$. The model can be described by using the curvature~$\kappa(s)$ of the curve~$\gamma(s)$, by the following \ac{ODE}
\begin{equation}
	\label{eq:model}
	\dot{x}=
	f(x, u)=
	\begin{bmatrix}
		\frac{v \cos(\beta)}{1-n \kappa(s)}\\
		v \sin(\beta)\\
		\frac{v}{l}\tan(\delta) - \frac{\kappa(s) v \cos(\beta)}{1-n \kappa(s)}\\
		\frac{1}{m}(F^\mathrm{d}-F^\mathrm{res}(v))\\
		r
	\end{bmatrix}.
\end{equation}
The function~$F^\mathrm{res}(v)=c_\mathrm{air}v^2+c_\mathrm{roll}\mathrm{sign}(v)$ models the air and rolling friction with constants~$c_\mathrm{air}$ and $c_\mathrm{roll}$.

\begin{figure}
        \vspace{2mm}
	\begin{center}
		\def\svgwidth{.4\textwidth}
\begingroup%
  \makeatletter%
  \providecommand\color[2][]{%
    \errmessage{(Inkscape) Color is used for the text in Inkscape, but the package 'color.sty' is not loaded}%
    \renewcommand\color[2][]{}%
  }%
  \providecommand\transparent[1]{%
    \errmessage{(Inkscape) Transparency is used (non-zero) for the text in Inkscape, but the package 'transparent.sty' is not loaded}%
    \renewcommand\transparent[1]{}%
  }%
  \providecommand\rotatebox[2]{#2}%
  \newcommand*\fsize{\dimexpr\f@size pt\relax}%
  \newcommand*\lineheight[1]{\fontsize{\fsize}{#1\fsize}\selectfont}%
  \ifx\svgwidth\undefined%
    \setlength{\unitlength}{265.51275539bp}%
    \ifx\svgscale\undefined%
      \relax%
    \else%
      \setlength{\unitlength}{\unitlength * \real{\svgscale}}%
    \fi%
  \else%
    \setlength{\unitlength}{\svgwidth}%
  \fi%
  \global\let\svgwidth\undefined%
  \global\let\svgscale\undefined%
  \makeatother%
  \begin{picture}(1,0.81116078)%
    \lineheight{1}%
    \setlength\tabcolsep{0pt}%
    \put(0,0){\includegraphics[width=\unitlength,page=1]{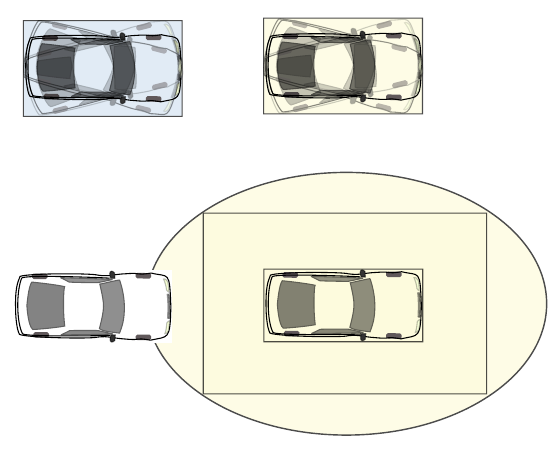}}%
    \put(0.24017614,0.54940322){\color[rgb]{0,0,0}\makebox(0,0)[lt]{\lineheight{1.25}\smash{\begin{tabular}[t]{l}$\setEgoCirc{}$\end{tabular}}}}%
    \put(0.71206303,0.55416376){\color[rgb]{0,0,0}\makebox(0,0)[lt]{\lineheight{1.25}\smash{\begin{tabular}[t]{l}$\setOpp{}$\end{tabular}}}}%
    \put(0.0699912,0.34270588){\color[rgb]{0,0,0}\makebox(0,0)[lt]{\lineheight{1.25}\smash{\begin{tabular}[t]{l}$[p_x,\;p_y]^\top$\end{tabular}}}}%
    \put(0.48582799,0.38264574){\color[rgb]{0,0,0}\makebox(0,0)[lt]{\lineheight{1.25}\smash{\begin{tabular}[t]{l}$\setOppExact{}$\end{tabular}}}}%
    \put(0,0){\includegraphics[width=\unitlength,page=2]{obstacle_sketch3.pdf}}%
    \put(0.52590901,0.04726711){\color[rgb]{0,0,0}\makebox(0,0)[lt]{\lineheight{1.25}\smash{\begin{tabular}[t]{l}$\setOppEll{}$\end{tabular}}}}%
    \put(0,0){\includegraphics[width=\unitlength,page=3]{obstacle_sketch3.pdf}}%
    \put(0.84950926,0.70903989){\color[rgb]{0,0,0}\makebox(0,0)[lt]{\lineheight{1.25}\smash{\begin{tabular}[t]{l}$\ub{\beta}$\end{tabular}}}}%
  \end{picture}%
\endgroup%

		\caption{Sketch of the considered \opp{} in Frenet coordinates. The occupied spaces for the ego vehicle~$\setEgoCirc{}$ and the \opp{}~$\setOpp{}$ are approximated by rectangles that contains all road-aligned heading angle configurations.
		By taking the Minkowski sum of both, which results in the inflated exact shape~$\setOppExact$, the ego vehicle shape can be considered as point with the position~$[p_x, \;p_y]^\top$. The set~$\setOppEll{}$ over-approximates the exact rectangular shape by an ellipse.}
		\label{fig:obstacle_sketch}
	\end{center}
\end{figure}

All possible configurations of vehicle shapes in the Frenet frame can be over-approximated by road-aligned rectangles.
Taking the Minkowski sum of the rectangular shapes for the ego and for an \opp{} yields an inflated rectangle that allows to consider the ego vehicle as a point mass~\cite{Schuurmans2023}, cf. Fig.~\ref{fig:obstacle_sketch}.
Notably, the accurate obstacle formulations including both vehicle configurations are computationally demanding, thus challenging to use in \nmpc{}.

Pivotal to this work is the accurate representation of rectangular obstacle avoidance constraints.
The over-approximated and inflated \opp{} shape has the length~$l$, the width~$w$, summarized in $\theta=[l,\; w]^\top$ and the \opp{} state is $z=[s^\mathrm{sv}\;,n^\mathrm{sv}\;,\beta^\mathrm{sv},\;v^\mathrm{sv},\;\delta^\mathrm{sv}]^\top$. 
With a scaling matrix 
\begin{equation}
	A(\theta)=\begin{bmatrix}
	\frac{2}{l}	& 0\\
	0	& \frac{2}{w}
	\end{bmatrix},
\end{equation}
and the projection matrix~$\selMatX\in\R^{2\times5}$ that selects the position states, the \opp{} shape can be normalized via the transformation 
$\nu : \R^2 \rightarrow \R^2$,
defined as
\begin{equation}
	\label{eq:nu}
	\nu(p;z,\theta) = A(\theta)(p-\selMatX z),
\end{equation}
which is linear in~$p$. With the normalized square denoted by 
\begin{align}
	\ball=\Bigl\{\gen\in\R^2 \Big| \norm{\gen}_\infty \leq 1 \Bigr\},
\end{align}
the obstacle can now be described as the following set
\begin{align}
	\label{eq:obstacle_set_exact}
	\setOppExact{}(z,\theta)=
	\Bigl\{	p \in \R^2 \Big|\nu(p;z,\theta) \in \ball \Bigr\}.
\end{align} 
A superscript~$j$ is used to refer to a particular \opp{}. Accordingly, the obstacle-free space with respect to a single obstacle described by the state~$z^j$ and parameters~$\theta^j$ can be written as
\begin{align}
	\label{eq:free_set}
	\begin{split}
	\setFreeOppExcat{}(z^j,\theta^j)=
	\Bigl\{	p \in \R^2 \Big|
	p \notin \setOppExact{}(z^j,\theta^j)\Bigr\}.
\end{split}
\end{align}
Besides obstacle avoidance constraints, further constraints can be expressed for the states by the admissible set~$\stateAdmiss{}$ and for the control inputs by the set~$\conrlAdmiss{}$.

In the following, the \emph{nominal} \nmpc{} formulation is introduced, which uses the exact obstacle formulation in the free set of~\eqref{eq:free_set}. A \ac{NLP} is formulated by direct multiple shooting~\cite{Bock1984} for a horizon of~$N$ steps. The model~\eqref{eq:model} is discretized with a step size~$\td{}$ 
to obtain the discrete-time dynamics~$x_{i+1}=F(x_i,u_i; \td)$. By setting a reference for states~$\refer{x}_i$ and controls~$\refer{u}_i=0$ in the \fcf{} with weights~$Q$ and~$R$, and using a projection matrix~$\selMatV\in\R^{1\times5}$ that selects the velocity state, the \ac{NLP} is 
	\begin{mini!}
		{\begin{subarray}{c}
				x_0, \ldots, x_N,\\
				u_0, \ldots, u_{N\!-\!1}
		\end{subarray}}			
		{\sum_{i=0}^{N-1} \norm{u_i}_{R}^2 \!+\! \norm{x_i-\refer{x}_{i}}_{Q}^2 \!+\! \norm{x_N-\refer{x}_{N}}_{Q_N}^2}
		{\label{eq:MPC}} 
		{} 
		\addConstraint{x_0}{= \hat{x}_0}{}
		\addConstraint{x_{i+1}}{= F(x_i,u_i; \td),}{ i\in \mathbb{I}(N-1)}
		\addConstraint{u_i}{\in \conrlAdmiss{},}{i\in \mathbb{I}(N-1)}
		\addConstraint{x_i}{\in \stateAdmiss{},}{i\in \mathbb{I}(N)}
		\addConstraint{\selMatV x_N }{= 0}{}
		\label{eq:nmpc_terminal_constraint}
		\addConstraint{\selMatX x_i}{\in \setFreeOppExcat(z_i^j,\theta_i^j),\; }{i\in \mathbb{I}(N)}\nonumber
		\addConstraint{}{}{j\in \mathbb{I}(M-1)}.
		\label{eq:nmpc_obstacle}
	\end{mini!}
The \ac{NLP}~\eqref{eq:MPC} is linearized at the previous solution to obtain a parametric \ac{QP}, which is solved within each iteration of the \ac{RTI} scheme~\cite{Diehl2005} after obtaining the state measurement~$\hat{x}_0$.
Note that the control admissible set also contains constraints for the lateral acceleration, cf.~\cite{Reiter2023a}. The ego vehicle is considered safe with zero velocity and no constraint violations. Hence, for simplicity, the equality constraint~\eqref{eq:nmpc_terminal_constraint} is used as a terminal safe set to obtain recursive feasibility.

\section{Progressive Smoothing in Real-time NMPC}
\label{sec:method}
In the following, the main contribution of the paper is introduced, an \ac{NMPC} scheme that replaces the highly non-smooth constraint~\eqref{eq:nmpc_obstacle} by a formulation that is successively smoothing the constraints along the prediction horizon. 

The good performance of the ellipsoidal constraint formulation~\cite{Reiter2023a} can be explained by favorable linearizations within \ac{SQP} iterations that are often used to implement \nmpc{}~\cite{Rawlings2017}.
The ellipsoidal \opp{} shapes are smooth and aligned with the road. Successive linearizations allow the shooting nodes to \emph{traverse} around the smooth shape of the collision region. 
We use higher-order norms that are progressively smoothed along the prediction horizon and referred to as \pnorm{}. At last prediction step~$N$, the \pnorm{} is equal to the ellipsoid. 

In the following, we use generalized coordinates~$\gen\in\R^n$ for introducing the smooth over-approximations of the unit hypercube. 
For the considered vehicle motion planning problem, we have $n=2$ and $\gen$ is obtained via the linear transformation in~\eqref{eq:nu}, i.e. $\gen=\nu(p;z,\theta)$. 

Formally, given two continuous functions $f(\gen):\R^n\rightarrow\R$ and $g(\gen):\R^n\rightarrow \R$, a homotopy map that depends on a homotopy parameter~$\alpha\in[\lb{\alpha},\ub{\alpha}]$, with $\ub{\alpha}\in\R \cup \{\infty\}$, is a continuous function $o(\gen,\alpha):\R^n\times [\lb{\alpha},\ub{\alpha}] \rightarrow \R$, with $o(\gen,\lb{\alpha})=f(\gen)$ and $o(\gen,\ub{\alpha})=g(\gen)$, for all $\gen\in\R^n$.
The concept of a homotopy map is used in the following to transition from a smooth constraint~$o(\gen,\lb{\alpha})$ to a tight constraint $o(\gen,\ub{\alpha})$. 

\begin{table}
	\centering
	\begin{tabular}{@{}lccccc@{}}
		\addlinespace
	  Property & 
        \begin{turn}{60} \pnorm{} \end{turn}& 
        \begin{turn}{60} \logsumexp{} \end{turn}& 
        \begin{turn}{60} \boltzmann{} \end{turn}&
        \begin{turn}{60} p-norm \end{turn}  & 
        \begin{turn}{60} \panos{} \end{turn}         \\
		\midrule
        Progressive Smoothing   &\checkmark & \checkmark & \checkmark& \xmark     & \xmark  \\
        Convexity              &\checkmark &\checkmark & \xmark & \checkmark & \checkmark  \\
        Over-approximation      & \checkmark & \checkmark & \checkmark& \checkmark & \checkmark \\
        Homogeneity     & \checkmark & \xmark & \xmark& \checkmark & \xmark \\
        Exact slack penalty     & \checkmark & \checkmark & \checkmark & \checkmark & \xmark  \\
		\bottomrule
	\end{tabular}
    \caption{Properties of the considered obstacle formulations.}
	\label{tab:comparison}
\end{table}
\emph{Convexity} of the \opp{} shape in $\gen$ is essential since it guarantees safe over-approximation within \ac{SQP}
iterates, cf.~\cite{Tran2012, Reiter2023a}. 
The \emph{tightening} property is crucial for recursive feasibility and defined as follows.
\begin{definition}
    A homotopy $o(\gen,\alpha):\R^n\times [\lb{\alpha},\ub{\alpha}] \rightarrow \R$ is monotonously \emph{tightening} with an increasing~$\alpha$, if for all $\alpha_2\geq\alpha_1$
	\begin{align}
		\begin{split}
			&\{\gen\in\R^n|o(\gen,\alpha_2)\leq1\} \subseteq \{\gen\in\R^n|o(\gen,\alpha_1)\leq1\}.
		\end{split}
	\end{align}
\end{definition}

The property of \emph{over-approximation} is used to describe whether for any value of the homotopy parameter~$\alpha \in [\lb{\alpha},\ub{\alpha}]$, the smooth shape is over-approximating the rectangular shape~$\ball$.
\begin{definition}
	A homotopy $o(\gen,\alpha):\R^n\times [\lb{\alpha}, \ub{\alpha}] \rightarrow \R$ is an \emph{over-approximation} of $\ball$, if for $\alpha \in [\lb{\alpha},\ub{\alpha}]$ it holds that
	\begin{align}
		\begin{split}
			&\ball \subseteq \{\gen\in\R^n|o(\gen,\alpha)\leq1\}.
		\end{split}
	\end{align}
\end{definition}

\begin{definition}
A homotopy $o(\gen,\alpha):\R^n\times \R^+ \rightarrow \R$ is tight w.r.t. $\ball$ if it is an over-approximation of $\ball$ and 
\begin{align}
\lim_{\alpha \rightarrow \ub{\alpha}} ~\{\gen\in\R^n|o(\gen,\alpha)\leq1\} = \ball.
\end{align}
\end{definition}

\subsection{\emph{\pnorm} Formulation}
The proposed \pnorm{} formulation is given via the homotopy $\obsFun^\mathrm{p}(\gen ;\alpha)$ defined as
\begin{equation}
	\obsFun^\mathrm{p}(\gen ;\alpha) = \Bigg(\frac{1}{n}\sum_{i=1}^{n}  \left|\gen_i\right|^{\alpha}\Bigg)^\frac{1}{\alpha},
\end{equation}
with homotopy parameter $\alpha \in [2, \infty)$.
Note that $\obsFun^\mathrm{p}(\gen ;\alpha)$ can be expressed as $\obsFun^\mathrm{p}(\gen ;\alpha) = \Vert n^{-\frac{1}{\alpha}} \gen \Vert_{\alpha}$ where $\Vert \cdot \Vert_{p}$ denotes the standard $p$-norm. 
For $\alpha = 2$, the set $\setOppPnorm(\alpha) = \{\gen\in\R^n|\obsFun^\mathrm{p}(\gen,\alpha)\leq1\}$ corresponds to a $n$-ball of radius $\sqrt{2}$. 
For $\alpha \rightarrow \infty$, we recover the unit square.

The following lemma shows that the \pnorm{} formulation is monotonously tightening and over-approximating, as well as convex in $\gen$. 

\begin{lemma}
The homotopy $\obsFun^\mathrm{p}(\gen ;\alpha)$  with $\alpha \in [2, \infty)$ and defining the sets $\setOppPnorm(\alpha) = \{\gen\in\R^n|\obsFun^\mathrm{p}(\gen,\alpha)\leq1\}$ has the following properties:
\begin{enumerate}
\item[(i)] The function $\obsFun^\mathrm{p}(\gen ;\alpha)$ is convex in $\gen$.
\item[(ii)] The sets $\setOppPnorm(\alpha)$ are over-approximations of $\ball$.
\item[(iii)] The sets $\setOppPnorm(\alpha)$ are monotonously tightening in $\alpha$.

\end{enumerate}
\end{lemma}
\begin{proof}
(i) Convexity in $\xi$ follows directly from convexity of the $\alpha$-norm for $\alpha\in [2, \infty)$.

(ii) Regard $\gen \in \R^n$ and let $\gen_{\max} = \max_i |\gen_i|$. 
We have 
\begin{align}
\obsFun^\mathrm{p}(\gen ;\alpha)
\leq
\left(\frac{1}{n}\sum_{i=1}^{n}  \left|\gen_{\max}\right|^{\alpha}\right)^\frac{1}{\alpha}
= 
\gen_{\max},
\end{align}
which shows $\obsFun^\mathrm{p}(\gen ;\alpha)
\leq \Vert\gen\Vert_\infty$ and thus $\ball \subseteq \setOppPnorm(\alpha)$. 

(iii) Let $\alpha_1 \!\leq \alpha_2$.
With $\Vert\gen\Vert_{\alpha_1}
\leq n^{\frac{1}{\alpha_1}-\frac{1}{\alpha_2}} \Vert\gen\Vert_{\alpha_2}$, we obtain
\begin{align}
\obsFun^\mathrm{p}(\gen ;\alpha_1)
& = n^{-\frac{1}{\alpha_1}} \Vert \gen \Vert_{\alpha_1}  \leq n^{-\frac{1}{\alpha_2}}  \Vert \gen \Vert_{\alpha_2} = \obsFun^\mathrm{p}(\gen ;\alpha_2).
\end{align}
\end{proof}


\subsection{Alternative Formulations}
\begin{figure}
    \vspace{1mm}
	\centering
	\includegraphics[trim={4mm 0 10mm 0},clip,scale=0.593]{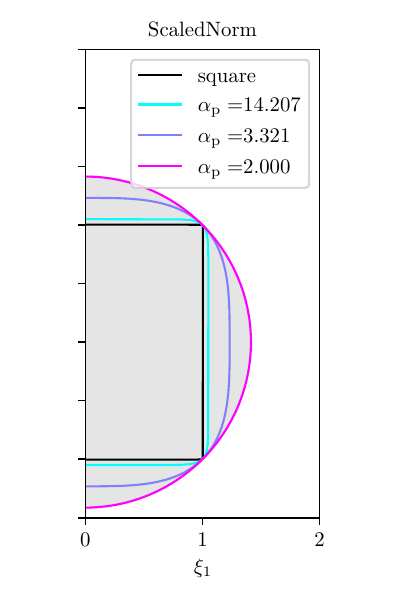}
	\includegraphics[trim={12mm 0 13mm 0},clip,scale=0.593]{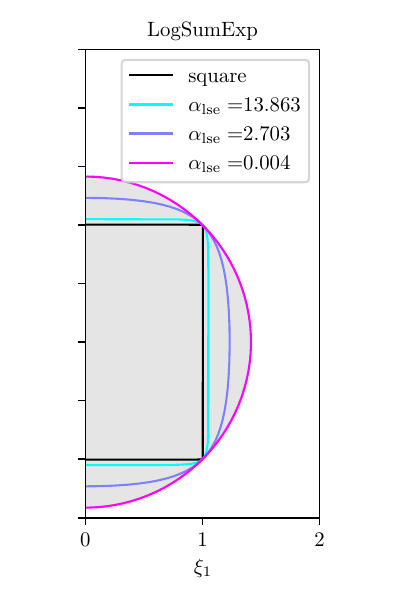}
	\includegraphics[trim={12mm 0 13mm 0},clip,scale=0.593]{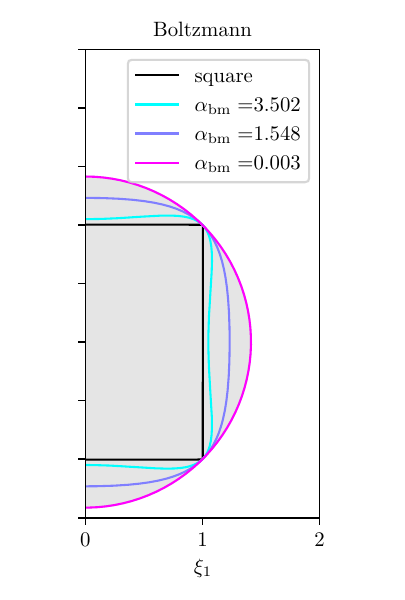}
	\caption{Obstacle shape smoothing in normalized coordinates for the set~$o(\gen,\alpha)=1$. The associated tightening parameters $\alpha_{\{\cdot\}}$ are smoothed from a square shape (black) to a circle (magenta). The values of~$\alpha_{\{\cdot\}}$ are chosen, such that the approximated widths and height are equal, measured at the axes. Notably, the non-convexity of the Boltzmann approximation can be seen for high values of~$\alpha_\mathrm{bm}$.}
	\label{fig:norm_comparisons}
\end{figure}

We compare the proposed progressively smoothing \pnorm{} formulation to four alternative constraint formulations:
(a) a higher order norm formulation ~\cite{Schimpe2020} that is constant along the prediction horizon; 
(b) the \panos{} formulation as introduced in ~\cite{Sathya2018};
(c) a progressively smoothing \logsumexp{} formulation;
(d) a progressively smoothing \boltzmann{} formulation.
Their properties are summarized in Tab.~\ref{tab:comparison}.
A visual comparison of the three progressively smoothing formulations -- the \pnorm{} formulation, the \logsumexp{} formulation, and the \boltzmann{} formulation -- is given in Fig.~\ref{fig:norm_comparisons}.

The \logsumexp{} formulation is defined via 
\begin{equation}
\obsFun^\mathrm{lse}(\gen;\alpha) = \eta_\mathrm{lse}(\alpha)\log\frac{1}{2n}
\sum_{i=1}^{n}\exp(\alpha \gen_i)+\exp(-\alpha \gen_i),
\end{equation}
with homotopy parameter~$\alpha \in (0, \infty)$ and normalization constant $\eta_\mathrm{lse}(\alpha)$ given as
\begin{equation}
\eta_\mathrm{lse}(\alpha)=\frac{1}{\log(\frac{1}{2}(\exp(\alpha)+\exp(-\alpha)))}.
\end{equation}
The corresponding sets $\setOppLog(\alpha) = \{\gen\in\R^2|\obsFun^\mathrm{lse}(\gen,\alpha)\leq1\}$ over-approximate the unit square.
For $\alpha \rightarrow 0$, we obtain $\setOppPnorm(2)$; and for $\alpha \rightarrow \infty$, we recover the unit square as illustrated in Fig.~\ref{fig:norm_comparisons}.
Furthermore, convexity of the \logsumexp{} function implies convexity of the sets  $\setOppLog(\alpha)$.

The \boltzmann{} formulation, which is based on the Boltzmann (also \emph{soft-max}) operator,  is given as
\begin{equation}
\obsFun^\mathrm{bm}(\gen ;\alpha)\!=\!\eta_\mathrm{bm}(\alpha) \frac{\sum_{i=1}^{n} \gen_i \exp(\alpha\gen_i) -\gen_i \exp(-\alpha\gen_i) }{\sum_{i=1}^{n}\exp(\alpha\gen_i) + \exp(-\alpha\gen_i)},
\end{equation}
with homotopy parameter $\alpha \in (0, \infty)$ and normalization constant $\eta_\mathrm{bm}(\alpha)$ defined as
\begin{equation}
\eta_\mathrm{bm}(\alpha)=\frac{\exp(\alpha) + \exp(-\alpha)}{\exp(\alpha) - \exp(-\alpha)}.
\end{equation}

As illustrated in Fig. \ref{fig:norm_comparisons}, the sets $\setOppBoltz(\alpha_\mathrm{bm}) = \{\gen\in\R^n|\obsFun^\mathrm{bm}(\gen,\alpha_\mathrm{bm})\leq1\}$ over-approximate the unit square. 
For $\alpha \rightarrow 0$, we recover the set $\setOppPnorm(2)$; and for $\alpha \rightarrow \infty$, we obtain the unit square $\ball$.
However, the \boltzmann{} function is nonconvex as illustrated by its nonconvex sublevel sets shown in Fig. \ref{fig:norm_comparisons}.

\subsection{Constraint Linearization and Homogeneity}

\begin{figure}
\vspace{2mm}
	\centering
	\includegraphics[trim={4mm 10mm 3mm 3mm},clip,scale=0.85]{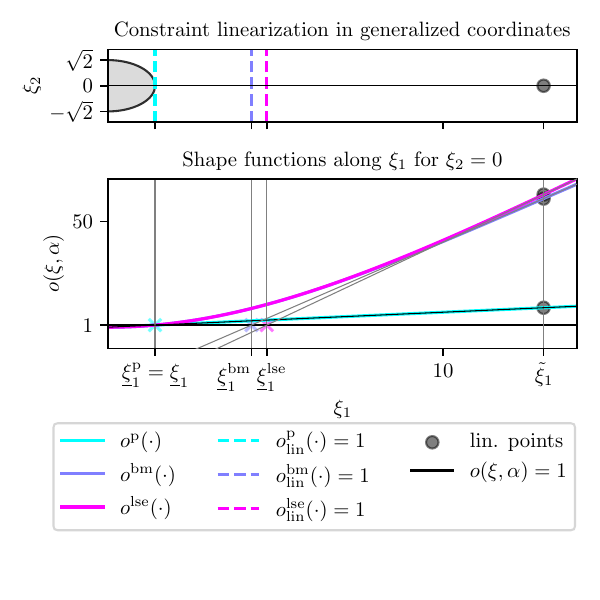}
	\caption{Linearized constraints~$o^{\{\cdot\}}_\mathrm{lin}(\cdot)\geq 1\Leftrightarrow\gen\geq\lb{\gen}^{\{\cdot\}}$ for the generalized coordinates~$\gen$ at an linearization point~$\tilde{\gen}=[\tilde{\gen}_1,0]^\top$ for the smoothest approximation, i.e., $\alpha_\mathrm{p}=2$ and $\alpha_\mathrm{lse}=\alpha_\mathrm{bm}\approx 0$. Related separating hyperplanes, i.e.,~$o^{\{\cdot\}}_\mathrm{lin}(\cdot)= 1$ are shown in the upper plot and the linearizations of the obstacle shape functions~$o^{\{\cdot\}}(\cdot)$ along~$\gen_1$ for $\gen_2=0$ are shown in the lower plot. Any p-norm is homogeneous, thus also for the \pnorm{} formulation it is true that $\underline{\gen}=\underline{\gen}^\mathrm{p}$ for any linearization point~$\tilde{\gen}$.}
	\label{fig:exat_liearization}
\end{figure}

During \ac{SQP} or \ac{RTI} iterations, the constraint function is linearized at the current iterate, here denoted by~$\tilde{\gen}$, in order to derive the QP sub-problem.
The following lemma shows that the linearization of the constraint is exact in the direction of the linearization point, i.e. the separating hyperplane obtained by linearizing the constraint is tight, if the \pnorm{} formulation is used.
This property, which is due to homogeneity of the norm, is not shared by the \boltzmann{} and \logsumexp{} formulation as illustrated in Fig.~\ref{fig:exat_liearization}.

\begin{lemma}
Let $\obsFun_\mathrm{lin}^p(\gen; \tilde{\gen}, \alpha)$ denote the linearization of $\obsFun^p(\gen; \alpha)$ at a linearization point $\tilde{\gen} \neq 0$. 
It holds that
\begin{align}
\obsFun_\mathrm{lin}^p(\gamma \tilde{\gen}; \tilde{\gen}, \alpha) = \obsFun^p(\gamma \tilde{\gen}; \alpha)
\end{align}
for all $\gamma \in [0, 1]$.
\end{lemma}

\begin{proof}
The partial derivative of $\obsFun^p(\gen; \alpha)$ is given as
\begin{align}
\frac{\partial \obsFun^p}{\partial \gen_i}(\gen; \alpha)
=
\frac{1}{n^{\frac{1}{\alpha}}}
\left(\frac{|\gen_i|}{\Vert \gen \Vert_p}\right)^{p-1}\mathrm{sign}(\xi_i),
\end{align}
from which we conclude that $\nabla_{\!\gen} \obsFun^p (\gen; \alpha) = \nabla_{\!\gen} \obsFun^p (\gamma\gen; \alpha)$ for any $\gamma \in [0, 1]$.
For $\gen = \gamma \tilde{\gen}$, we thus obtain
\begin{align}
\obsFun^p(\gen; \alpha)
& =
\obsFun^p(\tilde{\gen}; \alpha) + \!\!\int_{0}^1\! \nabla_{\!\gen} \obsFun^p (\tilde{\gen} + \tau (\gen - \tilde{\gen}); \alpha) (\gen - \tilde{\gen}) \, \mathrm{d}\tau
\nonumber \\
& =
\obsFun^p(\tilde{\gen}; \alpha) + \nabla_{\!\gen} \obsFun_\mathrm{lin}^p (\tilde{\gen}; \tilde{\gen}, \alpha)^{\top} (\gen - \tilde{\gen})
\nonumber \\
&= \obsFun_\mathrm{lin}^p(\gen; \tilde{\gen}, \alpha).
\end{align}
\end{proof}
\subsection{\nmpc{} Formulation using Progressive Smoothing}
For each of the tightening formulations, a scheduling function $\alpha_i^{\{\mathrm{p,lse,bm}\}}$ is used that parameterizes the tightening parameter~$\alpha$ according to the prediction time.
Consequently, for each \opp{}~$j$ and prediction step~$i$, the free set is defined as
\begin{align}
	\label{eq:obstalce_homotopies_final}
\begin{split}
	\freeSpace&_i^{\{\mathrm{p,lse,bm}\}}(z^j,\theta^j)=\\
	&\Bigl\{	p \in \R^2 \Big|
	\obsFun^{\{\mathrm{p,lse,bm}\}}(p;\alpha_i^{\{\mathrm{p,lse,bm}\}},z^j,\theta^j)\geq1\Bigr\},
\end{split}
\end{align}
and replaces the exact but non-smooth \opp{} constraint in~\eqref{eq:nmpc_obstacle}.
The parameters~$\alpha_i^{\{\mathrm{p,lse,bm}\}}$ parameterize the shape along the \nmpc{} prediction index~$i$ and yield the \emph{smoothest} over-approximation for~$i=\dimHor$ and the tightest for~$i=0$.

\begin{corollary}
With any of the three constraint homotopies, the resulting NMPC formulation satisfies recursive feasibility if the homotopy parameters $\alpha$ are chosen as a non-increasing sequence, i.e., $\alpha_i\geq\alpha_{i'}$ for $i\leq i'$.
\end{corollary}
\begin{proof}
This follows directly from the tightening property together with the terminal constraint $v_N = 0$.
\end{proof}

\subsection{Implementation using \ac{RTI} Algorithm}
In the \ac{RTI} algorithm for \nmpc{}, only one \ac{QP} is solved per time step, where the \ac{QP} is constructed by linearizing around the shifted solution guess from the previous time step. Due to high sampling rates of the controller and relatively slow parameter changes in the problem, the \rti{} solution is \emph{tracking} the optimum~\cite{Diehl2005} \emph{over time}.

The presented algorithm is particularly suited for~\ac{RTI}, since the shape parameters~$\alpha$ along the horizon remain constant throughout iterations. A contrary approach would be to solve several~\acp{QP} in one time step with an increasing shape parameter~$\alpha$ in each iteration, i.e., a homotopy in~$\alpha$, but equal for the whole horizon. The latter requires multiple \ac{QP} iterations per time step and is therefore computationally more expensive than the  \ac{RTI} algorithm. 
Additionally, the initial guess at the beginning of each new homotopy could be infeasible with respect to the smoothed constraints.

The homogeneity property of the \pnorm{} formulation is particularly beneficial for \ac{RTI} since even if the \ac{SQP} method did not fully converge, the linearization is exact. Thus, no over-approximation error due to the linearizations is made.

\section{Simulation Experiments}
\label{sec:experiments}
Using closed-loop \ac{NMPC} simulations with a single-track vehicle model, it is shown how the progressive smoothing of obstacle shapes outperforms the ellipsoidal formulation of~\cite{Reiter2023a}, the approach of~\cite{Sathya2018}, referred to as \panos{}, and using higher-order ellipsoids over the whole horizon~\cite{Schimpe2020}.

\begin{figure*}	
\vspace{2mm}
	\centering
	\includegraphics[scale=0.8,trim={5mm 0 0 15mm},clip]{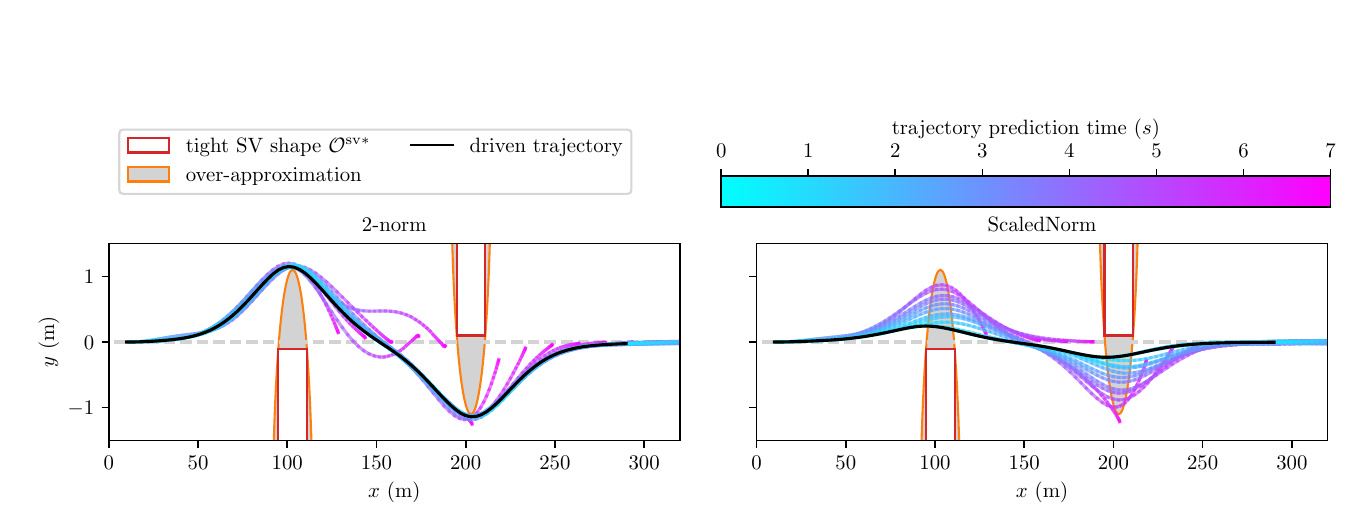}
	\caption{Comparison of two evasion maneuvers for the $2$-norm and the \pnorm{} formulation for a closed-loop simulation of $12s$. The $2$-norm formulation leads to the ego vehicle evading the two static obstacles conservatively due to the inflated shape of the~\opp{}. On the other hand, using the \pnorm{} formulation, the smooth transition of conservatively evading planned trajectories towards a tight driven trajectory (black) is visible.}
	\label{fig:trajectories}
\end{figure*}

As an illustrative example, an evasion of two static obstacles is simulated for the \pnorm{} and the $2$-norm, where the planned trajectories and the actually driven trajectories are compared, cf., Fig.~\ref{fig:trajectories}. 
Furthermore, three randomized scenario types are simulated while evaluating key performance indicators relevant for the overtaking performance and the computation time for each approach.
\subsection{Setup}
The experiments involve two different simulation setups, where each experiment highlights a representative key performance indicator.
Both experiments involve a randomized road, a simulated ego vehicle and one slower preceding \opp{}, formulated as single-track models as in~\cite{Reiter2023a}. The simulated states are~$\hat{x}_i$ at the simulation time~$t=i\td{}$. One single simulation is executed for~$t_\mathrm{sim}=15$ seconds, corresponding to~$N_\mathrm{sim}=\frac{t_\mathrm{sim}}{\td{}}$ simulation steps, which generously allows overtaking. Parameters for both vehicles are taken from the \texttt{devbot~2.0} specification, which can be found in~\cite{Reiter2021}, and which correspond to a full-sized race car that was used in the real-world competition~\texttt{Roborace}~\cite{Roborace2020}. For the \opp{}, the chassis width and length, as well as the maximum speed was randomized to capture different shapes. In the following, the individual simulation setups are described and associated parameters are shown in Tab.~\ref{tab:experiments}. 
\begin{table}
	\centering
	\begin{tabular}{@{}lcc@{}}
		\addlinespace
		\toprule
		 Parameter & \multicolumn{2}{c}{Values} \\
		  & Experiment 1 & Experiment 2  \\
		\midrule
		curvature      			& \multicolumn{2}{c}{$[0.01, 0.06] \frac{1}{\mathrm{m}}$} \\
		road width      		&\multicolumn{2}{c}{$10\mathrm{m}$}\\
		\midrule
		\opp{} set velocity     &  \multicolumn{2}{c}{$[0,5]\frac{\mathrm{m}}{\mathrm{s}}$}\\ 
		\opp{} width      		&   \multicolumn{2}{c}{$[1.5,4]\mathrm{m}$}\\
		\opp{} length      		&  $[4,14]\mathrm{m}$ &  $[2,10]\mathrm{m}$\\
		\opp{} start pos.		&  \multicolumn{2}{c}{$[50,120]\mathrm{m}$}\\
		\midrule
		ego set velocity		& \multicolumn{2}{c}{$[7,15]\frac{\mathrm{m}}{\mathrm{s}}$} \\
		ego weight~$w_n$		&$5$& $50$\\
		ego start pos. 			&\multicolumn{2}{c}{$[0,10]\mathrm{m}$} \\
		\bottomrule
	\end{tabular}
	\caption{Parameters used for simulation. Randomized parameters are uniformly sampled from the given interval.}
	\label{tab:experiments}
\end{table}

\subsubsection*{Experiment 1 - Lateral Distance}
\begin{figure}
	\begin{center}
		\def\svgwidth{.5\textwidth}
		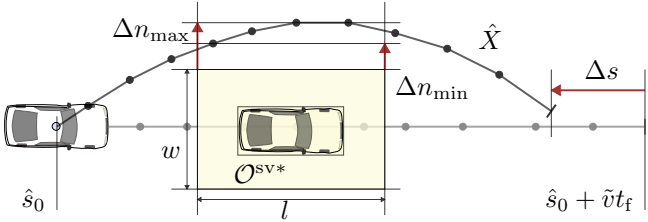
		\caption{Sketch of the evaluated performance measures of the simulated trajectory~$\hat{X}$, including the minimal and maximal overtaking distance~$\Delta n_\mathrm{min}$ and~$\Delta n_\mathrm{max}$, respectively, and the distance~$\Delta s$ to the maximum position that could be reached without \opps{}. }
		\label{fig:obstacle_sketch_measures}
	\end{center}
\end{figure}

The first experiment evaluates the maximum lateral distance
\begin{equation}
    \Delta n_\mathrm{max}=\max_{i\in \mathbb{J}} \Big|\hat{n}_i - n^{\mathrm{SV}}_i - \frac{w}{2}\Big|,
\end{equation}
and minimum lateral distance
\begin{equation}
    \Delta n_\mathrm{min}=\min_{i\in \mathbb{J}} \Big|\hat{n}_i - n^{\mathrm{SV}}_i - \frac{w}{2}\Big|,
\end{equation}
where 
\begin{equation}
\mathbb{J}=\bigg\{i\in \mathbb{I}(N_\mathrm{sim})\bigg|s^{\mathrm{SV}}_i - \frac{l}{2}\leq \hat{s}_i \leq s^{\mathrm{SV}}_i+ \frac{l}{2}\bigg\},
\end{equation}
while overtaking, cf. Fig.~\ref{fig:obstacle_sketch_measures}. It reveals that the actual driven trajectory is influenced by the over-approximations and verifies whether the constraint was violated.
\begin{figure*}
\vspace{2mm}
	\centering
	\includegraphics[scale=0.9]{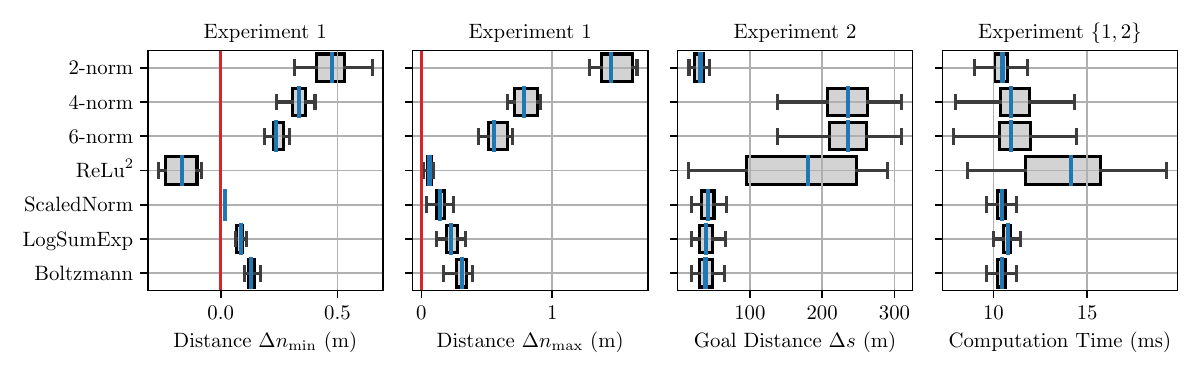}
	\caption{Box-plot performance evaluation of the proposed approaches in different randomized closed-loop experiments. In \emph{Experiment~1}, the maximum and minimum lateral distance~$\Delta n_\mathrm{max}$ and ~$\Delta n_\mathrm{min}$ to an \opp{} while overtaking is evaluated. A negative distance indicates an unsafe constraint violation (red line) as observed with the \panos{} formulation. In \emph{Experiment~2}, the goal distance~$\Delta s$, i.e., the difference of the maximum reachable position to the actual position, after overtaking at the final simulation time~$t_\mathrm{sim}$ is evaluated. The computation time was evaluated for all experiments.}
	\label{fig:evaluation}
\end{figure*}
\subsubsection*{Experiment 2 - Center Line Tracking}
\label{sec:evaluation_exp3}
In many real-world applications, the driving cost is partly specified by tracking a certain reference line, which involves a higher cost to avoid cutting corners. If this cost is high or if the slower preceding \opp{} has a larger width, a local minimum of the optimization problem~\eqref{eq:MPC} can be created behind the \opp{}. Non-smooth \opp{} shape representations tend to promote this local minimum, which is evaluated in the second experiment, by setting the center line tracking cost~$w_n$ higher. The performance measure $\Delta s$ is the difference of the maximum reachable final position and the actual final position, written as
\begin{equation}
	\label{eq:eval_delta_s}
	\Delta s=\hat{s}_0  + \refer{v}\dimT - \hat{s}_{N_\mathrm{sim}-1},
\end{equation}
with the positions $\hat{s}_i$ and the set velocity~$\refer{v}$ at the final simulation time~$t_\mathrm{sim}$.

\subsubsection*{Nonlinear Model Predictive Controller}
The \nmpc{} is formulated using the \ac{NLP}~\eqref{eq:MPC} with the different \opp{} approximations according to~\eqref{eq:obstalce_homotopies_final} and model parameters according to the \texttt{devbot}, cf.~\cite{Roborace2020, Reiter2021}.

The shape parameters~$\alpha_i^{\{\cdot\}}$ at index~$i$ are determined implicitly, by defining the width and the height~$\refer{d}_i$ of the square obstacle shape at the axes of the auxiliary coordinates~$\xi$, and implicitly solving the equation
\begin{equation}
    o^{\{\cdot\}}\left([0,\;\refer{d}_i]^\top, \alpha_i^{\{\cdot\}}\right)=1,
\end{equation}
offline. The width~$\refer{d}$ can vary between $\refer{d}=1$, which would correspond to the exact rectangle and $\refer{d}=\sqrt{2}$, corresponding to the circle or the $2$-norm. In the experiments, we linearly change the over-approximated width from~$\refer{d}_0=1.005$ to~$\refer{d}_N=\sqrt{2}$ which implicitly defines the shape parameters~$\alpha_i^{\{\cdot\}}$ for all steps~$i$. 

Fig.~\ref{fig:lin_prog} shows the related shapes for indexes~$i\in\{0,\frac{N}{2},N\}$. By defining the shape based on the width, the progressive smoothing formulations are nearly equal in terms of their over-approximated area and allow for fair comparisons between the different approaches.

\subsubsection*{Implementation details}
To solve the \nmpc{} problem~\eqref{eq:MPC}, the solver~\texttt{acados}~\cite{Acados2021}, together with the \ac{QP} solver \texttt{HPIPM}~\cite{Frison2020} is used. We use \ac{RTI}, without condensing, an explicit RK4 integrator and a Gauss-Newton Hessian approximation. For the prediction, $N=70$ shooting nodes are used with a discretization time of~$\td{}=0.1$ seconds which corresponds to a prediction horizon of~$7$~seconds.

\subsection{Evaluation}
In Fig.~\ref{fig:trajectories}, actual driven and planned trajectories of an overtaking maneuver are shown in the Cartesian coordinates. It can be clearly seen in the left plot that the ellipsoidal formulation leads to conservative behavior, in which the ego vehicle performs strong lateral swaying maneuvers to avoid any potential collision with the \opps.

In Fig.~\ref{fig:evaluation}, the key performance measures of the experiments are shown. As already indicated by Fig.~\ref{fig:trajectories}, the maximum lateral distance~$\Delta n_\mathrm{max}$ is largest for the ellipsoidal formulation. Increased order ellipsoids, such as the $4$-norm and the $6$-norm, yield superior results for the maximum lateral distance, similarly as the proposed progressive smoothing and \panos{} formulations.

When evaluating the minimum distance~$\Delta n_\mathrm{min}$ while overtaking, the \panos{} formulation reveals a disadvantageous property, i.e., even though the obstacle slack variables are chosen as exact penalties (L1 penalization) for all formulations with equal weights, the constraints are violated. This issue was mentioned in~\cite{Sathya2018} and circumvented by an increased over-approximation.
All other methods respect the constraints exactly and are therefore considered safe.

With an increased center line tracking cost, the constant higher order ellipsoids and the \panos{} formulation repeatedly get stuck in the local minima behind the preceding \opp{}, limiting its performance and basically getting blocked behind. This problem is the main reason why other works suggest not using constant higher-order norms and rather use the $2$-norm~\cite{Reiter2023a, Reiter2023b}. For any overtaking objectives, this problem can be limiting. 
The simulation results empirically show that the proposed approaches are less likely to get stuck in a local minimum, and therefore lead to an improved performance for this particular case study.

In the final plot of Fig.~\ref{fig:evaluation}, the computation times among all experiments are shown and verify that the computation time is similar for each of the considered \ac{NMPC} formulations, despite an average increase of the median computation time by $27.2\%$ and the maximum computation time by $60.9\%$ using the \panos{} formulation compared to the \pnorm{}.

The \pnorm{} formulation has a good performance in all evaluations, taking into account the constraint violation of the \panos{} formulation. The reason for the \pnorm{} performing better than the \logsumexp{} and \boltzmann{} may be the exact constraint linearization related to the norm-function. Despite the good performance of also the \boltzmann{} formulation, it can not be guaranteed that the linearized constraint safely over-approximates the obstacle shape~\cite{Reiter2023a} due to its non-convex shape.

\section{Conclusions}
\label{sec:conclusion}
A novel progressive smoothing scheme was presented for obstacle avoidance in \nmpc{}. The presented \pnorm{} approach, as well as the alternative formulations \boltzmann{} and \logsumexp{} outperform the benchmarks, including fixed higher-order ellipsoids~\cite{Schimpe2020} and the formulation used in~\cite{Sathya2018}. The performance achieved by the \pnorm{} is superior to all others in these particular experiments, which is likely due to the advantageous linearizations of the norm-function. The alternative \boltzmann{} formulation has the drawback of non-convexity. The \pnorm{} formulation also has desirable theoretical properties, such as convexity, tightening, homogeneity, exact slack penalty and over-approximation.


\section*{Acknowledgment}
Rudolf Reiter, Katrin Baumgärtner and Moritz Diehl were supported by EU via ELO-X 953348, by DFG via Research Unit FOR 2401, project 424107692 and 525018088 and by BMWK via 03EI4057A and 03EN3054B.


\bibliographystyle{IEEEtran}
\bibliography{lib}

\end{document}